\newcommand{\suma}[1]{\sum_{{#1} \in \mathbb{Z}}}
\newcommand{\op}[1]{\hat {#1}}
\newcommand{\openone}{\leavevmode\hbox{\small1\normalsize\kern-.33em1}}
\newcommand{\Tr}{\mathop{\mathrm{Tr}}\nolimits}

\documentclass[final,1p,times,eqsecnum]{elsarticle}

\usepackage{graphicx,amssymb,amsmath,bm,subeqnarray,color,ul
em,hyperref}

\journal{Annals of Physics}

\begin{document}

\begin{frontmatter}

\title{Orbital angular momentum in phase space}

\author[ucm]{I. Rigas} 
\author[ucm]{L. L. S\'{a}nchez-Soto}
\author[udg]{A. B. Klimov}
~\author[upol]{J. \v{R}eh\'{a}\v{c}ek}
\author[upol]{Z. Hradil}

\address[ucm]{Departamento de \'Optica, 
Facultad de F\'{\i}sica, Universidad Complutense, 
28040~Madrid, Spain}

\address[udg]{Departamento de F\'{\i}sica,
Universidad de Guadalajara, 
44420~Guadalajara, Jalisco, Mexico}

\address[upol]{Department of Optics,
Palack\'{y} University, 17. listopadu 12,
746 01 Olomouc, Czech Republic}

\date{\today}

\begin{abstract}
  A comprehensive theory of the Weyl-Wigner formalism for the
  canonical pair angle-angular momentum is presented.  Special
  attention is paid to the problems linked to rotational periodicity and
  angular-momentum discreteness.
\end{abstract}

\end{frontmatter}

\section{Introduction}

Phase-space methods were introduced in the very early times of quantum
mechanics to avoid some of the troubles arising in the abstract
Hilbert-space formulation. The pioneering works of
Weyl~\cite{Weyl:1928}, Wigner~\cite{Wigner:1932uq}, and
Moyal~\cite{Moyal:1949fk} paved thus the way to formally representing
quantum mechanics as a statistical theory on phase
space~\cite{Stratonovich:1956kx,Agarwal:1970vn,Berezin:1975ys,Agarwal:1981zr,
Bertrand:1987ly,Varilly:1989bh,Brif:1998qf,Benedict:1999dq}.

In this approach one looks for a mapping relating operators (in
Hilbert space) to functions (in phase space). In particular,
quasiprobability distributions are the functions connected with the
density operator~\cite{Tatarskii:1983uq,Balazs:1984cr,Hillery:1984oq,
Lee:1995tg, Schroek:1996fv,Schleich:2001hc,QMPS:2005}.  Usually, the
relevant theoretical tools are illustrated with continuous variables
(such as Cartesian position and momentum), for which much physical
knowledge has been inferred. Likewise, the increasing role of qudits
in modern quantum information has fuelled a lot of interest in
consistently extending these achievements to the case of a discrete
phase space~\cite{Wootters:1987kl,Galetti:1988ff,Galetti:1992pi,
Wootters:2004fu,Gibbons:2004ye,Vourdas:2004qo,Klimov:2006ee,
Vourdas:2007tw,Bjork:2008ab,Klimov:2009bk,Durt:2010pd}.

However, other physical problems call for different topologies of the
phase
space~\cite{Laidlaw:1971fk,Schulman:1971vn,Dowker:1972kx,Acerbi:1993fr}.
Specifically, in this paper we focus on the discrete cylinder
$\mathcal{S}_1 \times \mathbb{Z}$ ($\mathcal{S}_1$ denotes here the
unit circle and $\mathbb{Z}$ the integers), which is associated to
the canonical pair angle-angular momentum.  A number of systems, such
as molecular rotations, electron wave packets, Hall fluids, and light
fields, to cite only a few examples, can be described in terms of this
geometry~\cite{Allen:2003,Vortices:2008xw}.  In quantum optics, it is
of primary importance to deal with the orbital angular momentum of
twisted photons~\cite{Molina:2007kn,Franke-Arnold:2008sw}, which have
been proposed for numerous
applications~\cite{Vaziri:2002,Langford:2004ve}.
 
The proper definition of angles in quantum mechanics has a long
history and requires more care than perhaps might be
expected~\cite{Carruthers:1968qq,Lynch:1995cq,Perinova:1998xq,Luis:2000fk}.
Since the conjugate angular momentum has an unbounded spectrum (that
includes positive and negative integers), it is, in principle,
possible to introduce a \textit{bona fide} angle
operator. Periodicity, however, brings out subtleties that have
triggered long and heated discussions.  We think that a phase-space
treatment of this variable may shed light on the origin of all those
problems.

A pioneer attempt in that direction was made by
Mukunda~\cite{Mukunda:1979uq,Mukunda:2005kx}, who for the first time
constructed a Wigner function on the discrete cylinder.  This work was
subsequently reelaborated and developed in a variety of directions by
other
authors~\cite{Bizarro:1994vn,Vourdas:1996ys,Nieto:1998cr,Ruzzi:2002rt,Zhang:2003zr,Kakazu:2006fr}.
Coherent states for this pair have also attracted much
attention~\cite{Kowalski:1996mz,Gonzalez:1998gf,Ohnuki:1993ul,
  Hall:2002pd,Ruzzi:2006dq,Rigas:2010kx}. These accomplishments, as
well as some other related questions, such as the associated minimum
uncertainty states, are reviewed in great detail by
Kastrup~\cite{Kastrup:2006cr}.

In spite of this considerable progress, there is still the need, in our view, 
for a comprehensive geometrical approach to this problem, which allows us to 
integrate in a structured way all the previous knowledge, while providing
the right tools to attack recent and attractive problems, such as, e.g.,
the tomography of these systems~\cite{Rigas:2008nx}. The purpose of this
paper is precisely to provide such a description.

\section{Phase space for quantum continuous variables}
\label{sec:pscv}

In this section we recall the structures needed to set up a
phase-space description of Cartesian quantum mechanics.  This will
facilitate comparison with the cylinder later on. For simplicity, we
choose one degree of freedom, so the associated phase space is the
plane $\mathbb{R}^2$.

The relevant observables are the Hermitian coordinate and momentum
operators $\op{q}$ and $\op{p}$, with canonical commutation relation
(with $\hbar = 1$ throughout)
\begin{equation}
  \label{eq:HWcom}
  [\op{q}, \op{p}] = i \, \op{\openone} \, ,
\end{equation}
so that they are the generators of the Heisenberg-Weyl
algebra~\cite{Binz:2008oq}.  Ubiquitous and profound, this algebra has
become the hallmark of noncommutativity in quantum theory. 

To avoid technical problems with the unboundedness of $\op{q}$ and
$\op{p}$, it is convenient to work with their unitary counterparts
\begin{equation}
  \op{U} (q) = \exp (-i q \, \op{p}) \, ,
  \qquad 
  \op{V} (p) = \exp (i p \, \op{q}) \, ,
  \label{UV}
\end{equation}
whose action in the bases of eigenvectors of position and momentum is
\begin{equation}
  \label{eq:actbas}
  \op{U} (q^{\prime} ) | q \rangle = | q + q^{\prime} \rangle  \, ,
  \qquad 
  \op{V} (p^{\prime} ) | p \rangle = | p + p^{\prime} \rangle \, ,
\end{equation}
so they represent displacements along the corresponding coordinate
axes. The commutation relations are then expressed in the Weyl
form~\cite{Galindo:1991fk}
\begin{equation}
  \op{V}(p) \op{U}(q) = e^{iqp} \, \op{U}(q) \op{V}(p) \, .  
  \label{eq:Weyl}
\end{equation}
Their infinitesimal version immediately gives (\ref{eq:HWcom}), but
(\ref{eq:Weyl}) is more useful in many instances.

In terms of $\op{U}$ and $\op{V}$ a general displacement operator can
be introduced as
\begin{equation}
  \label{eq:HWDisp1}
  \op{D} (q,p) = \op{U} (p)  \op{V}(q)  e^{- i q p/2} = 
  \exp[i(p \op{q} -  q \op{p})] \, ,
\end{equation}
with the parameters $(q,p)$ labeling phase-space points.  These
operators form a complete trace-orthonormal set (in the continuum
sense) in the space of operators acting on $\mathcal{H}$ (the Hilbert
space of square integrable functions on $\mathbb{R}$):
\begin{equation}
  \label{eq:HWDispOrtho}
  \Tr [ \op{D} (q, p) \, \op{D}^\dagger (q^\prime, p^\prime) ] = 
  2 \pi \, \delta (q - q^\prime) \delta (p - p^\prime)  \,  .
\end{equation}
Note that $\op{D}^\dagger (q, p) = \op{D} (-q, - p)$, while
$\op{D}(0,0) = \op{\openone}$.  In addition, they obey the simple
composition law
\begin{equation}
  \label{eq:com} 
  \op{D} (q, p) \, \op{D} (q^{\prime}, p^{\prime} ) = 
  \exp[i (q p^{\prime} - p q^{\prime})/2] \, 
  \op{D} (q + q^{\prime}, p + p^{\prime} ) \, .
\end{equation}

We can also work with the Fourier transform of $\op{D} (q, p)$
\begin{equation}
  \label{eq:HWkernelDef}
  \op{w} (q, p) = \frac{1}{(2\pi)^2} \int_{\mathbb{R}^2} 
  \exp[-i(p q^\prime -q p^\prime)] \, \op{D} (q^\prime, p^\prime) \, 
  dq^\prime dp^\prime \, ,
\end{equation}
which is called a Stratonovich-Weyl
quantizer~\cite{Stratonovich:1956kx}.  One can check that the
operators $\op{w} (q, p)$ are also a complete trace-orthonormal set
that transforms properly under displacements
\begin{equation}
  \label{eq:HWKernelDisp}
  \op{w} (q, p) = \op{D} (q,p) \,\op{w} (0, 0) \,
  \op{D}^\dagger (q, p) \, ,
\end{equation}
where
\begin{equation}
  \label{eq:HWPar1}
  \op{w}(0,0)=\int_{\mathbb{R}^{2}} \op{D}(q, p) \, dq dp = 2  \op{P}
  \, ,
\end{equation}
and
\begin{equation}
  \label{eq:2}
  \op{P} = \int_{\mathbb{R}} | q \rangle \langle - q | \, dq = 
  \int_{\mathbb{R}} | p \rangle \langle - p | \, d p 
\end{equation}
is the parity operator.

Let $\op{A}$ be an arbitrary (Hilbert-Schmidt) operator acting on
$\mathcal{H}$.  Using the Stratonovich-Weyl quantizer we can associate
to $\op{A}$ a tempered distribution $a(q, p)$ on $\mathbb{R}^{2}$
representing the action of the corresponding dynamical variable in
phase space. In fact, this is known as the Wigner-Weyl map and reads
as
\begin{equation}
  \label{eq:Asy}
  a(q, p)  =   \Tr [ \op{A} \,\op{w}(q,p) ] \, . 
\end{equation}
The function $a (q, p)$ is the symbol of the operator $\op{A}$. Conversely, we
can reconstruct the operator from its symbol through
\begin{equation}
  \label{eq:WigWeyl}
  \op{A} = \frac{1}{(2\pi)^{2}} \int_{\mathbb{R}^{2}} a (q, p) \, 
  \op{w} (q, p ) \, dq dp \, .
\end{equation}

The symbols form an associative algebra endowed with a noncommutative
(star) product inherited by the operator product. Let the functions
$a(q,p)$ and $b(q, p)$ correspond to the operators $\op{A}$ and
$\op{B}$, respectively. If we denote by $(a \star b) (q, p)$ the
function corresponding to $\op{A} \op{B}$, some simple manipulations
lead us to~\cite{Schroek:1996fv}
\begin{equation}
  \label{eq:starint}
  (a \star b) (q, p) = \frac{4}{(2 \pi)^{2}} \int_{\mathbb{R}^{2}} 
  \int_{\mathbb{R}^{2}} a(q + q^{\prime}, p + p^{\prime}) \,
  \exp[ 2 i ( q^{\prime} p^{\prime \prime} - q^{\prime \prime} p^{\prime}) ] \,
  b(q + q^{\prime \prime}, p + p^{\prime \prime} ) \, dq^{\prime} dp^{\prime} 
  dq^{\prime \prime} dp^{\prime \prime} \, . 
\end{equation}
Expanding the functions $a$ and $b$ in a formal Taylor series at the 
point $(q, p) \in \mathbb{R}^{2}$ we obtain
\begin{equation}
  \label{eq:stardiff}
  (a \star b) (q, p) = a (q,p) \, \exp \left (- \frac{i}{2}
    \overleftrightarrow{\mathcal{P}} \right ) \, b (q, p) \, ,
\end{equation}
where $\overleftrightarrow{\mathcal{P}}$ is the Poisson operator
\begin{equation}
  \label{eq:Poiss}
  \overleftrightarrow{\mathcal{P}} = 
  \frac{\overleftarrow{\partial}}{\partial q}
  \frac{\overrightarrow{\partial}}{\partial p} -
  \frac{\overleftarrow{\partial}}{\partial p}
  \frac{\overrightarrow{\partial}}{\partial q} \, ,
\end{equation}
in terms of which we can define the Moyal bracket~\cite{Moyal:1949fk}
\begin{equation}
  \label{eq:Moyalbrac}
  \{ a, b \}_{M} = \frac{1}{i} (a \star b - b \star a) \, ,
\end{equation}
which is the phase-space counterpart of the commutator in the
Hilbert-space formulation. Consequently, the time evolution of 
$a(q,p)$ can be expressed as
\begin{equation}
  \label{eq:evoleq}
  \frac{d a}{dt} = \{h, a \}_{M} \, ,
\end{equation}
where $h$ is  the symbol of the Hamiltonian.

In this context, the Wigner function is nothing but the symbol of the
density matrix $\op{\varrho}$. Therefore, 
\begin{eqnarray}
  \label{eq:Wigcan}
  &  W_{\op{\varrho}}(q, p)  =   \Tr [ \op{\varrho} \,\op{w}(q,p) ] \, , & \nonumber \\ 
  & &  \label{eq:HWWignerDef} \\
  & \op{\varrho}   =   \displaystyle 
  \frac{1}{(2\pi)^{2}} \int_{\mathbb{R}^{2}} \op{w}(q,p) W_{\op{\varrho}}(q,p) \, dq dp \, . &
  \nonumber
\end{eqnarray}
For a pure state $| \Psi \rangle$, it can  be represented as
\begin{equation}
  \label{eq:1}
  W_{\op{\varrho}} (q, p) = \frac{1}{2\pi}  \int_{\mathbb{R}}  \exp (i p q^{\prime} ) \,
  \Psi (q - q^{\prime}/2) \, \Psi^{\ast} (q + q^{\prime}/2 )  \, dq^{\prime} \, ,
\end{equation}
which is, perhaps, the most convenient form for actual calculations.
According to (\ref{eq:evoleq}),  the Liouville-von Neumann
equation  in phase space is 
\begin{equation}
  \label{eq:evoleqW}
  \frac{d W_{\op{\varrho}}}{dt} = \{h, W_{\op{\varrho}} \}_{M} \, .
\end{equation}
When applying this equation in practical cases, one needs the
following mapping for the action of the basic variables $\op{q}$ and
$\op{p}$ on $\op{\varrho}$:
\begin{eqnarray}
  \label{eq:correspXP}
  \begin{array}{rcl}
    \op{q} \,\op{\varrho} &\mapsto & \displaystyle 
    \left ( q -   \frac{i}{2} \frac{\partial}{\partial p} \right) W_{\op{\varrho}}(q,p)  \, , \\ 
    & &\\
    \op{\varrho} \,\op{q}&\mapsto &  \displaystyle 
    \left ( q +  \frac{i}{2} \frac{\partial}{\partial p} \right) W_{\op{\varrho}}(q,p)
    \, ,
  \end{array}
  &
  \begin{array}{rcl}
    \op{p}\,\op{\varrho} &\mapsto & \displaystyle 
    \left ( p +  \frac{i}{2}\frac{\partial}{\partial q} \right) W_{\op{\varrho}}(q,p)
    \, , \\
    & &\\
    \op{\varrho} \,\op{p}&\mapsto & \displaystyle 
    \left( p -  \frac{i}{2}\frac{\partial}{\partial q}\right) W_{\op{\varrho}}(q,p) \, .
  \end{array}
\end{eqnarray}

The Wigner function defined in (\ref{eq:HWWignerDef}) fulfills all the
basic properties required for any good probabilistic description.
First, due to the Hermiticity of $\op{w} (q,p)$, it is real for
Hermitian operators. Second, on integrating $W (q, p)$ over the lines
$q_{\phi} = q \cos \phi + p \sin \phi$, the probability distributions of
the rotated quadratures $q_{\phi}$ are reproduced
\begin{equation}
  \label{eq:rotquadmad}
  \int_{\mathbb{R}^{2}} W_{\op{\varrho}} (q, p) \, \delta (q - q_{\phi} ) \,dq dp =
  \langle q_{\phi} | \op{\varrho} | q_{\phi} \rangle \, .
\end{equation}
In particular, the probability distributions for the canonical
variables can be obtained as the marginals
\begin{equation}
  \label{eq:HWProps2}
  \int_\mathbb{R}  W_{\op{\varrho}} (q, p) \, dp =   
  \langle q | \op{\varrho} | q \rangle   \, ,
  \qquad
  \int_\mathbb{R} W_{\op{\varrho}}(q, p) \, dq = 
  \langle p | \op{\varrho} | p \rangle  \, .
\end{equation}
Third, $W_{\op{\varrho}}(q, p)$ is translationally covariant, which
means that for the displaced state $\op{\varrho}^\prime =
\op{D}(q^{\prime}, p^{\prime})$ $\op{\varrho} \, \op{D}^\dagger
(q^{\prime}, p^{\prime})$, one has
\begin{equation}
  \label{eq:HWProps3}
  W_{\op{\varrho}^\prime} (q, p) = W_{\op{\varrho}} (q-q^{\prime}, p-p^{\prime}) \, ,
\end{equation}
so that it follows displacements rigidly without changing its form,
reflecting the fact that physics should not depend on a certain choice
of the origin. The same holds true for any linear canonical transformation.

Finally, the overlap of two density operators is proportional to the
integral of the associated Wigner functions:
\begin{equation}
  \label{eq:HWProps4}
  \Tr ( \op{\varrho} \,\op{\varrho}^{\prime} ) \propto
  \int_{\mathbb{R}^2} W_{\op{\varrho}} (q, p)  W_{\op{\varrho}^{\prime}} (q, p)  \, dq dp \, .
\end{equation}
This property (known as traciality) offers practical advantages,
since it allows one to predict the statistics of any outcome, once the
Wigner function of the measured state is known.

However, the Wigner function can take on negative values, a property
which distinguishes it from a true probability distribution.  Indeed,
this negativity is associated with the existence of quantum
interference, which itself may be identified as a signal of
nonclassical behavior~\cite{Kenfack:2004lw}.  The
characterization of quantum states that are classical, in the sense of
giving rise to nonnegative Wigner functions, is a topic of undoubted
interest. Among pure states, it was proven by
Hudson~\cite{Hudson:1974kb} that the only states that have
nonnegative Wigner functions are Gaussian
states~\cite{Janssen:1984kx,Lieb:1990yq}.

Coherent states are closely linked with the notion of Gaussian states.
The displacements constitute a basic ingredient for their definition:
indeed, if we choose a fixed normalized reference state $ |
\Psi_{0}\rangle $, we have~\cite{Perelomov:1986kl}
\begin{equation}
  | q, p \rangle = \op{D} ( q, p) \, | \Psi_{0} \rangle \, ,  
  \label{eq:defCS}
\end{equation}
so they are parametrized by phase-space points. These states have a
number of remarkable properties inherited from those of $\op{D} (q,
p)$. In particular, $\op{D} (q, p)$ transforms any coherent state in
another coherent state:
\begin{equation}
  \op{D} ( q^{\prime},  p^{\prime} ) \, | q, p \rangle
  = \exp[i (q^{\prime} p - p^{\prime} q )/2] \,
  | q + q^{\prime}, p + p^{\prime} \rangle \, .
  \label{eq:comcoh}
\end{equation}
The standard choice for the fiducial vector $| \Psi_{0} \rangle$ is
the vacuum $|0 \rangle $. This has quite a number of relevant
properties, but the one we want to stress for what follows is that $| 0 
\rangle $ is an eigenstate of the Fourier transform (as they are all
the Fock states)~\cite{Mehta:1987tg,Ruzzi:2006ai}. In fact, over this 
apparently trivial property  rests a huge amount of physical knowledge.   
So, $|\Psi_{0} \rangle$ is taken as the Gaussian
\begin{equation}
  \Psi_{0} (q) =
  \frac{1}{\pi^{1/4}} \, \exp ( - q^{2}/2) \, ,
  \label{eq:Gauss}
\end{equation}
in appropriate units. In addition, this wave function, as any coherent
state, represents a minimum uncertainty state, namely
\begin{equation}
  ( \Delta q )^{2} \, ( \Delta p)^{2} = \frac{1}{4} \, ,  
  \label{eq:MUS}
\end{equation}
where $(\Delta q)^{2}$ and $(\Delta p)^{2}$ are the corresponding
variances.

As a last illustration of the role played by the displacement
operators, we note that their completeness allows us to write for any
observable $\op{A}$
\begin{equation}
  \label{eq:Dtom}
  \op{A} = \frac{1}{2 \pi} \int_{\mathbb{R}^{2}}
  \Tr[ \op{A} \op{D}^{\dagger} (q, p ) ] \, \op{D} (q, p ) \,
  dq dp \, .
\end{equation}
Evaluating the trace in terms of the set of eigenvectors of the
rotated quadratures $\op{q}_{\phi}$ we obtain~\cite{Paris:2004,Lvovsky:2009dp}
\begin{equation}
  \label{eq:tompol}
  W_{\op{\varrho}} (q, p) = \frac{1}{2 \pi} \int_{0}^{\pi} d\phi 
  \int_{-\infty}^{\infty}  p (q_{\phi}, \phi) \,  K (q \cos \phi + p \sin \phi -
  q_{\phi})  \, dq_{\phi} d\phi  \, ,
\end{equation} 
where $K(x)$ is the kernel
\begin{equation}
  \label{eq:Ker}
  K (x) = \frac{1}{2} \lim_{\eta \downarrow 0} \mathrm{Re}
  \frac{1}{(x + i \eta)^{2}} \, .
\end{equation}
The tomograms $ p (q_{\phi}, \phi) = \langle q_{\phi} | \op{\varrho} | q_{\phi}
  \rangle$ are measured in a homodyne detector. This shows that these
rotated quadratures provide a complete quorum to reconstruct the
density operator and hence the Wigner function.

\section{Phase space for angle-angular momentum}
\label{sec:huds-theor-angle}

\subsection{Geometrical properties of the discrete cylinder}
\label{sec:angle}

In this Section we trace the changes required when we replace the
Cartesian position by a periodic angular position $\phi \in
\mathcal{S}_{1}$. For definiteness, we take the window $(- \pi, \pi )$
in further considerations. The canonical conjugate variable, denoted
now as $L$, is the component of the angular momentum along the axis
orthogonal to the rotation plane.  While classically a point particle
is necessarily located at a single value of the angle $\phi$, the
corresponding quantum wave function is an object extended around
$\mathcal{S}_1$ and can be directly affected by the nontrivial
topology.

One may be tempted to think that angular position should stand in the
same relationship to angular momentum as ordinary position stands to
linear momentum. This would prompt to use the commutation relation
$[\op{\phi}, \op{L} ] = i \, \op{\openone}$ and interpret the angle
operator as multiplication by $\phi$, while $\op{L}$ is the
differential operator $\op{L} = - i \partial_{\phi}$. Nevertheless,
the use of this operator may entail many pitfalls for the unwary and
needs a very subtle 
analysis~\cite{Carruthers:1968qq,Emch:1972,Uffink:1990}.

As stressed in the Introduction, a long discussion about the
properties of this angle operator turns out to be unnecessary to
settle a phase-space description. Indeed, let us denote by $| \ell
\rangle$ the  basis of angular-momentum eigenstates. In
principle, the corresponding eigenvalues can be arbitrary. However,
when some invariance principles
hold~\cite{Kowalski:1996mz,Kastrup:2006cr}, we can restrict our
attention to integer values. In this case, the states
\begin{equation}
  \label{phi_states} 
  |\phi \rangle = \frac{1}{\sqrt{2 \pi}}
  \suma{\ell}  e^{- i \ell \phi} | \ell \rangle \, ,
\end{equation}
constitute a basis of states with well-defined angle, so that $
\langle \phi | \phi^\prime \rangle = \delta_{2\pi} (\phi - \phi^\prime
)$, where $\delta_{2\pi}$ represents the periodic delta function (or
Dirac comb) of period $2 \pi$~\cite{Bracewell:1999lr}, and they allow
for a resolution of the identity.  Denoting the wave function
components in both bases as $\Psi_{\ell} = \langle \ell | \Psi \rangle
$ and $\Psi(\phi) = \langle \phi | \Psi \rangle$, we have
\begin{equation}
  \label{Fourier}
  \Psi ( \phi ) =   \frac{1}{\sqrt{2 \pi}} 
  \suma{\ell} e^{i \ell \phi} \,  \Psi_{\ell} \, ,
  \qquad
  \Psi_{\ell}  =  \int_{- \pi}^{\pi} \! \! \!
  e^{- i \ell \phi}  \,  \Psi(\phi)   \,  d\phi \, , 
\end{equation}
which translates the Fourier relationship inherent to any canonical
pair.

To proceed further, we restrict ourselves to the exponentiated version
of the canonical pair
\begin{equation}
  \op{U} (\phi ) = \exp (-i \phi \, \op{L}) \, ,
  \qquad 
  \op{V} ( \ell ) = \exp (i \ell \, \op{\phi}) \, ,
  \label{UVAM}
\end{equation}
which, in addition, are experimentally feasible
operations~\cite{Hradil:2006pl,Rehacek:2008ss}. We have the action
\begin{equation}
  \op{U} (\phi^{\prime} ) | \phi \rangle =  | \phi + \phi^{\prime} \rangle \, ,
  \qquad 
  \op{V} ( \ell^{\prime} ) | \ell \rangle =  | \ell + \ell^{\prime} \rangle \, ,
\end{equation}
so that they represent displacements along the coordinate axes of the
cylinder $\mathcal{S}_1 \times \mathbb{Z}$. Throughout, the angle
addition and subtraction must be understood modulo $2 \pi$. 
In  dealing with this pair, it is customary also to take as the
fundamental angular variable the complex exponential of the angle 
\begin{equation}
  \label{eq:defE}
  \op{E}= \exp (- i \op{\phi}) \, ,
\end{equation}
instead of the angle itself.

The Weyl form for this pair reads as
\begin{equation}
  \op{V} (\ell) \op{U}( \phi ) = e^{i\ell \phi} \, 
  \op{U} (\phi) \op{V}( \ell) \, , 
  \label{eq:WeylAM}
\end{equation}
and following the ideas of Section~\ref{sec:pscv}, the displacement
operators can be introduced as
\begin{equation}
  \label{eq:Displace1}
  \op{D} (\ell, \phi) = \exp[ i \alpha (\ell,\phi)]  \, 
  \op {V} (\ell ) \op{U} ( \phi ) \, ,
\end{equation}
where $ \alpha(\ell,\phi)$ is a phase. The reasonable condition
$\op{D}^\dag (\ell,\phi) = \op{D} (-\ell,-\phi) $ imposes 
\begin{equation}
  \label{eq:PhaseConditionAlpha}
  \exp[ i \alpha(\ell, \phi) + i \alpha (-\ell, -\phi) ] = 
  \exp (-i \ell \phi ) \, .
\end{equation}
We cannot rewrite now equation~(\ref{eq:Displace1}) as an
entangled exponential.

These displacement operators form a complete trace-orthonormal set
\begin{equation}
  \label{eq:DispOrtho}
  \Tr [ \op{D} ( \ell , \phi )  \, \op{D}^\dagger (\ell^\prime, \phi^\prime ) ] 
  =  2 \pi \, \delta_{\ell \ell^\prime} \, 
 \delta_{2\pi}(\phi - \phi^\prime) \, , 
\end{equation}
whose resemblance with the relation~(\ref{eq:HWDispOrtho}) is evident.
One can also introduce the Stratonovich-Weyl quantizer as
\begin{equation}
  \label{eq:WigKerDef1}
  \op{w} (\ell, \phi) = 
  \frac{1}{(2\pi)^2}  \suma{\ell^\prime} \int_{- \pi}^{\pi}   
  \exp[-i ( \ell^\prime \phi - \ell \phi^\prime)] \,
  \op{D} (\ell^\prime, \phi^\prime) \, d\phi^\prime \, ,
\end{equation}
so that it fulfills the covariance condition
  \begin{equation}
    \label{eq:CovarExplicit}
    \op{w} (\ell ,\phi ) = \op{D} (\ell ,\phi ) \, \op{w} (0,0)
    \,\op{D}^\dag (\ell ,\phi ) \, .
  \end{equation}
Note carefully that $\op{w} (0, 0)$ cannot be identified now with the
parity operator on the cylinder
\begin{equation}
  \label{eq:Parcyl}
  \op{P} = \sum_{\ell \in \mathbb{Z}} | \ell \rangle \langle - \ell |  =
\int_{- \pi}^{\pi} | \phi \rangle \langle - \phi | \, ∂d\phi  \, ,
\end{equation}
as it is the case for continuous variables. This is due to the fact
that $\op{D} (\ell ,\phi ) \, \op{P} \,\op{D}^\dag (\ell ,\phi )$ do
not constitute an operator basis. One gets a basis if one supplements
$\op{P}$ by $\op{E} \op{P}$ (or $\op{P} \op{E}$), $\op{E}$ being the
exponential of the angle (\ref{eq:defE}).
 
The symbol of an operator $\op{A}$ turns out to be
\begin{equation}
  a (\ell, \phi) = \Tr [ \op{A} \,\op{w} ( \ell,\phi) ] \, ,
\end{equation}
while the inversion is given by
\begin{equation}
  \op{A} = 2\pi  \suma{\ell} \int_{- \pi}^{\pi} 
  a(\ell, \phi) \, \op{w}(\ell,\phi)  \, d\phi \, . 
\end{equation}
As explained by Pleba\'nski and coworkers~\cite{Plebanski:2000fk}, this
Wigner-Weyl correspondence on the cylinder directly distinguishes
between the ``classical'' phase space $\mathcal{S}_{1} \times
\mathbb{R}$ from the ``quantum'' phase space $\mathcal{S}_{1} \times
\mathbb{Z} $: only the former admits a formalism based on the use of
symbols.

We can also look for the form of the star product. In close analogy
with equation~\eqref{eq:starint}, we have the integral form
\begin{equation}
  ( a \star b) (\ell,\phi)  =   \displaystyle 
  \frac{1}{2\pi} \suma{\ell^{\prime}, \ell^{\prime \prime} }
  \int_{-\pi}^\pi   a(\ell + \ell^{\prime}, \phi + \phi^{\prime} /2) \, 
  \exp[i(\ell^{\prime \prime} \phi^{\prime} -  
  \ell^{\prime}  \phi^{\prime \prime} ) ]  \, 
  b(\ell + \ell^{\prime \prime}, \phi + \phi^{\prime \prime} /2) \, 
  d\phi^{\prime} d\phi^{\prime \prime} \, . 
\end{equation}
Expanding the functions $a$ and $b$ in a formal Taylor series at
$(\ell,\phi)$, this result can also be rewritten in a ``differential''
form [compare with equation (\ref{eq:stardiff})]
\begin{equation}
  \label{eq:starE2diff}
  (a \star b) (\ell ,\phi) = a  (\ell,\phi) \, 
  \exp\left ( -\frac{i}{2}   \overleftrightarrow{\mathbb{P}} \right ) \,
  b (\ell, \phi) \, ,
\end{equation}
where $\overleftrightarrow{\mathbb{P}}$ is the counterpart of the
Poisson operator (\ref{eq:Poiss}) on the cylinder
\begin{equation}
  \overleftrightarrow{\mathbb{P}} = 
  \overleftarrow{\delta}_\ell 
  \frac{\overrightarrow{\partial}}{\partial \phi} -
  \frac{\overleftarrow{\partial}}{\partial
    \phi} \overrightarrow{\delta}_\ell \, . 
\end{equation}
Here $\delta_\ell$ is some sort of a continuous interpolation between
discrete translations, and is defined through
\begin{equation}
  \label{eq:deltaEll}
  \exp( \lambda \delta_\ell )   f (\ell) =
  \frac{1}{2\pi}\int_{-\pi}^\pi 
  \exp[ i (\ell + \lambda) \phi] \,\tilde f(\phi) \, d\phi \, ,
\end{equation}
$\tilde f (\phi)$ being the Fourier transform of the function
$f(\ell)$.

With this star product, we can introduce a Moyal bracket as in
equation (\ref{eq:Moyalbrac}) and the time evolution of dynamical
variables is also given by (\ref{eq:evoleq}).  

To conclude, some remarks concerning the phase $\alpha (\ell,
\phi)$ in (\ref{eq:Displace1}) are in order. Although many of the
results in this paper are independent of $\alpha (\ell, \phi)$, a
judicious choice of this factor may greatly facilitate calculations.
The periodicity in $\phi$ requires $\op{D} (\ell, \phi) = \op{D}
(\ell, \phi +  2\pi)$ [apart from
(\ref{eq:PhaseConditionAlpha})]. Obviously, there are many admissible
functions. For example, quadratic definitions like $\exp [i \alpha
(\ell,\phi)] = \exp[-i \ell (\ell +1) \phi /2]$ or $\exp[i\ell (\ell -
1)\phi /2]$ are legitimate, but we discard them because they introduce
a lot of difficulties when computing practical Wigner functions.

Another possibility  is 
\begin{equation}
  \label{eq:AlternativePhases}
 \exp[ i \alpha (\ell ,\phi)]  =  \exp (-i \ell \phi /2) \, \exp \{ i \phi \, 
 [1 -  (-1)^\ell ] / 4 \} \, , 
\end{equation}
leading to the Wigner kernel 
\begin{equation}
  \label{eq:AlternativeKernel}
  \op{w} (\ell , \phi )  = \frac{1}{2\pi} \op{D} (\ell  ,\phi) \,
  ( \op{P}  + \op{E}  \, \op{P } ) \, \op{D}^\dag (\ell ,\phi) \, ,   
\end{equation}
which, as discussed before, contains the complete set $\op{P}$ and
$\op{E} \op{P}$. In addition, this option brings the advantage of 
very easy calculations, especially for states given in the angular momentum
representation. However, in this case the term $\op{w}(0,0) \propto
\op{P} +  \op{E} \, \op{P } $ in Eq.~(\ref{eq:AlternativeKernel})  is not
parity invariant. Since the phase space should not have a ``preferred
direction'', we will also discard this definition. 

We finally fix the phase $\alpha$ as 
\begin{equation}
  \label{eq:AlphaExplicit}
 \exp [i \alpha (\ell,\phi ) ]  = \exp (-i \ell \phi /2 ) \, ,
\end{equation}
 which may lead to slightly more involved calculations than
 (\ref{eq:AlternativePhases}), but has the clear advantage of a
 parity-invariant kernel. We stress that this definition of  $\alpha$ 
is related to the choice of the $2 \pi$ window for $\phi$, so it
is exclusively valid for $\phi \in [-\pi, \pi]$.

\subsection{Wigner function on the cylinder}
\label{sec:wign-funct-discr}

Since the Wigner function is the symbol associated with the density
matrix, we have the  mapping 
\begin{eqnarray}
&  W_{\op{\varrho}} (\ell, \phi) = \Tr [ \op{\varrho} \,\op{w} ( \ell,\phi) ] \, , & 
\nonumber \\ 
& & \label{eq:WigFunDef1} \\
& \displaystyle
\op{\varrho} = 2\pi  \suma{\ell} \int_{- \pi}^{\pi} 
\op{w}(\ell,\phi) \, W_{\op{\varrho}}(\ell,\phi) \, d\phi \, . & 
\nonumber
\end{eqnarray}
To obtain an operational form of this Wigner function, we note that
with the phase definition (\ref{eq:AlphaExplicit}), 
the quantizer kernel (\ref{eq:WigKerDef1}) becomes
\begin{eqnarray}
  \label{eq:ExplicitKernel1}
  \op{w} (\ell,\phi) &  =  &  \displaystyle
  \frac{1}{(2\pi)^2} \suma{\ell^\prime,\ell^{\prime \prime}} \int_{- \pi}^{\pi}  
  \exp[i ( \ell^{\prime}/2 - \ell^{\prime \prime} ) \phi^{\prime}]  
  \exp[i (\ell \phi^{\prime} - \ell^{\prime} \phi) ] \,
  |\ell^{\prime \prime} \rangle 
  \langle \ell^{\prime \prime} - \ell^{\prime}| \,  d\phi^\prime \, . 
\end{eqnarray}
Because of the contribution with a half-integer exponent, to sum over
$\ell^{\prime}$ one has to split this sum in two different parts
according $\ell^{\prime}$ is either even or odd.  After some
manipulations, one gets
\begin{eqnarray}
  \label{eq:ExplictKernel3} 
  W_{\op{\varrho}} (\ell,\phi) & = & \displaystyle 
  \frac{1}{2\pi} \suma{\ell^\prime}  \exp(-2 i \ell^\prime \phi ) \,
  \langle \ell - \ell^\prime | \,\op{\varrho}\ |\ell + \ell^\prime \rangle  
  \nonumber  \\ 
  & + &  \displaystyle 
  \frac{1}{2\pi^2} \suma{\ell^{\prime},\ell^{\prime \prime}} 
  \frac{(-1)^{\ell^{\prime \prime}}}{\ell^{\prime \prime} + 1/2} 
  \exp[-i  (2 \ell^{\prime} + 1)  \phi]  \,
  \langle\ell +\ell^{\prime \prime}-\ell^{\prime}| \op{\varrho} 
  |\ell + \ell^{\prime \prime}+\ell^{\prime}+1\rangle \, . 
\end{eqnarray}
This expression contains exclusively angular-momentum matrix elements
of the density operator and it is involved in practical computations
due to the second line, which presents a logarithmically slow
convergence due to the fact that for odd $\ell$, the phase
(\ref{eq:AlphaExplicit}) as a function of $\phi$ has a discontinuity
at $\phi = \pm \pi$. Incidentally, this second line would be greatly
simplified by employing the alternative kernel
(\ref{eq:AlternativeKernel}).

Sometimes, it is preferable to work in the
angle representation, which can be obtained by introducing a
resolution of the unity in equation~(\ref{eq:WigKerDef1}). In this way, we
get
\begin{equation}
  \label{eq:WignerAngle}
  W_{\op{\varrho}}(\ell,\phi) = \frac{1}{2\pi} \int_{- \pi}^{\pi}  
  \exp (i \ell \phi^{\prime} )
  \langle \phi - \phi^{\prime}/2| \op{\varrho}\,|\phi + \phi^{\prime}/2\rangle \,
  d\phi^{\prime} \, ,
\end{equation}
which coincides with the form introduced in the pioneering work by
Mukunda~\cite{Mukunda:1979uq}. This Wigner function reproduces 
the proper marginal distributions
\begin{equation}
  \label{eq:ELMargin}
  \suma{\ell} W_{\op{\varrho}} (\ell,\phi) = 
  \langle \phi | \op{\varrho} | \phi \rangle \, ,
  \qquad
  \int_{-\pi}^{\pi} W_{\op{\varrho}} (\ell,\phi) \, d\phi = 
  \langle \ell|\op{\varrho} | \ell \rangle \, ,
\end{equation}
and it is explicitly covariant under displacements on the
cylinder. This confirms the good properties of this approach.

Note, in closing, that the time evolution of this Wigner function is
given by a Liouville-von Neumann equation analogous
to (\ref{eq:evoleqW}). Now, in practical calculations we shall
need the following map, which can be easily checked
\begin{equation}
  \label{eq:correspondence}
  \begin{array}{rcl}
    \begin{array}{rcl}
      \op{L} \, \op{\varrho} & \mapsto & \displaystyle 
 \frac{1}{2\pi} 
 \left ( \ell -\frac{i}{2} \frac{\partial}{\partial \phi}\right) W_{\op{\varrho}}
 (\ell,\phi)\, , \\
& & \\
 \op{\varrho} \,\op{L} & \mapsto &  \displaystyle 
 \frac{1}{2\pi}
\left(\ell + \frac{i}{2}\frac{\partial}{\partial \phi}\right) W_{\op{\varrho}} (\ell,\phi)\,,
    \end{array} &
    \begin{array}{rcl}
      \op{E} \,\op{\varrho} & \mapsto & \displaystyle
\frac{1}{2\pi} 
  e^{-i\phi} \,   e^{\frac{1}{2}\delta_\ell} \, W_{\op{\varrho}}(\ell,\phi) \, ,\\
&&\\
 \op{\varrho} \,\op{E} & \mapsto & \displaystyle
 \frac{1}{2\pi} \,e^{-i\phi} \,  e^{-\frac{1}{2}\delta_\ell} \, W_{\op{\varrho}}(\ell,\phi)\,.
    \end{array}
  \end{array}
\end{equation}
Again, we have taken as the fundamental angular variable the complex
exponential of  the angle instead of the angle itself.

\section{Examples}

To further appreciate these ideas, we present a few relevant examples.
For an angular momentum eigenstate $| \ell_0 \rangle$ and for an angle
eigenstate $|\phi_0 \rangle$, one has
\begin{equation}
  \label{eq:ExampleOAMstaleEll}
  W_{| \ell_0 \rangle} (\ell,\phi) = \frac{1}{2\pi} 
  \delta_{\ell, \ell_0} \, , 
  \qquad
  W_{| \phi_0 \rangle} (\ell, \phi) = \frac{1}{2\pi} \,
  \delta_{2\pi}(\phi-\phi_0) \, . 
\end{equation}
In the first case, the Wigner function is flat in $\phi$ and the
integral over the whole phase space gives the unity, reflecting the
normalization of $|\ell_0\rangle$, while in the second, it is flat in
the conjugate variable $\ell$, and thus, the integral over the whole
phase space diverges, which is a consequence of the fact that the
state $|\phi_0\rangle$ is not normalizable.

Next, we consider coherent states. The standard approach
due to Perelomov~\cite{Perelomov:1986kl} does not work for the
cylinder. However, much as in  (\ref{eq:defCS}), we can  introduce 
them as
\begin{equation}
  \label{eq:CScyl}
  | \ell, \phi \rangle = \op{D} (\ell, \phi ) \, | \Psi_{0} \rangle \, .
\end{equation}
The choice of the fiducial vector is not so obvious. We
take~\cite{Ruzzi:2006dq}
\begin{equation}
  \label{eq:fid}
  \Psi_{0} (\phi ) = \frac{1}{\sqrt{2 \pi}}  
  \frac{\vartheta_3 \left ( \frac{\phi}{2}  \big | \frac{1}{e^{1/2}} \right )}
  {\sqrt{\vartheta_3 \left ( 0  \big | \frac{1}{e} \right )}} \, ,
\end{equation}
where $\vartheta_3$ denotes the third Jacobi theta
function~\cite{Mumford:1983}, which incidentally plays the role of the
Gaussian in circular statistics.  In this way, the states $|\ell, \phi
\rangle$ inherit the properties of $\op{D} (\ell, \phi)$ and they turn
out to be equivalent to the ones defined in \cite{Kowalski:1996mz} by
an eigenvalue equation or in \cite{Gonzalez:1998gf} via a Zak
transform.

The Wigner function for the coherent state $|\ell_0, \phi_0\rangle$ splits as
$ W_{|\ell_0,\phi_0 \rangle} (\ell,\phi) =  W^{(+)}_{|\ell_0,\phi_0\rangle} (\ell, \phi) +  
  W^{(-)}_{|\ell_0,\phi_0\rangle} (\ell, \phi)$. The ``even'' part turns out to be
\begin{equation}
  W^{(+)}_{|\ell_0,\phi_0\rangle} (\ell,\phi) =  
  \frac{1}{2 \pi \vartheta_3 \left (0 \big |\frac{1}{e} \right )} 
  e^{-(\ell -\ell_0)^2} 
  \vartheta_3 \left(\phi-\phi_0 \Big|  \frac{1}{e} \right) \, .
\end{equation}
This seems a sensible result, since it is a discrete Gaussian in the
variable $\ell$, and the equivalent for a Gaussian for the continuous
angle $\phi$ it. However, the ``odd'' contribution spoils this simple
picture: 
\begin{equation}
  \nonumber
  W^{(-)}_{|\ell_0,\phi_0\rangle} (\ell,\phi) =
  \frac{\exp [ i (\phi-\phi_0)- 1/2]}{2\pi^2 \vartheta_3
    \left ( 0  \big | \frac{1}{e} \right )}
  \vartheta_3\left( \phi- \phi_0 +i/2 \Big|\frac{1}{e} \right) \, 
  \suma{\ell^\prime} (-1)^{\ell^\prime - \ell + \ell_0} 
  \frac{\exp(-\ell^\prime {}^2 - \ell^\prime)}{\ell^\prime + \ell_0 - \ell + 1/2} \, .
  \label{eq:ExampleWoddCoh}  
\end{equation}

\begin{figure}[t]
 \centerline{\includegraphics[width=0.90\columnwidth]{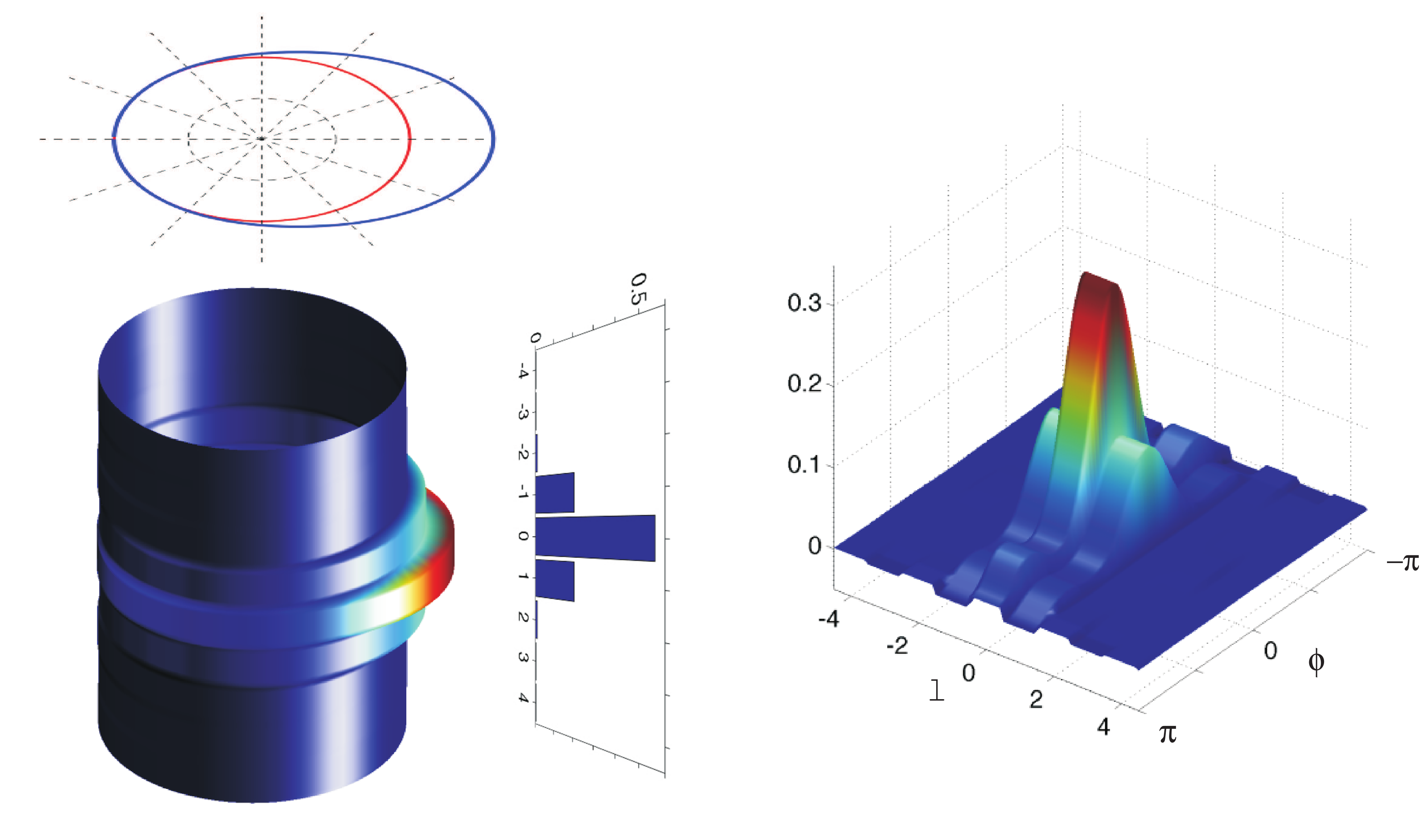}}
     \caption{Plot of the Wigner function for a coherent state with
    $\ell_0 =0$ and $\phi_0 = 0$. The cylinder extends vertically from
    $\ell = -4 $ to $\ell = +4$. The two corresponding marginal
    distributions are shown. In the right panel we show the unwrapped
    version of the Wigner function.}
  \label{fig:CoherentStandard}
\end{figure}

In figure~\ref{fig:CoherentStandard}, the Wigner function for the
coherent state $| \ell_0 = 0, \phi_0 = 0\rangle$ is plotted.  A
pronounced peak at $\phi=0$ for $\ell=0$ and slightly smaller ones for
$\ell=\pm 1$ can be observed.  The associated marginal distributions
are also plotted. They are strictly positive, as correspond to true
probability distributions.  For quantitative comparisons, however,
sometimes it may be convenient to ``cut'' this cylindrical plot along
a line $\phi$=constant and unwrap it. This is also shown in
figure~\ref{fig:CoherentStandard}.

A closer look at this figure reveals also a remarkable fact: for
values close to $\phi= \pm \pi$ and $\ell= \pm 1$, the Wigner function
takes negative values.  Actually, a numeric analysis suggests the
existence of negativities close to $\phi = \pm \pi$ for any odd value
of $\ell$.

As our last example, we address the superposition
\begin{equation}
  \label{eq:SuperDef}
  | \Psi \rangle = \frac{1}{\sqrt{2}} 
  (|\ell_1 \rangle   + e^{i\phi_0} | \ell_2 \rangle )
\end{equation}
of two angular-momentum eigenstates with a relative phase $e^{i
  \phi_0}$.  The Wigner function splits again; now the ``even'' part
reads as
\begin{equation}
  \label{eq:ExampleSuperEven}
  W_{  | \Psi \rangle}^{(+)} (\ell, \phi)   =  
  \frac{1}{4\pi}  \{ \delta_{\ell, \ell_1} + \delta_{\ell, \ell_2} +
  2 \delta_{\ell_ 1+\ell_2, 2\ell} \, \cos[\phi_0 + (\ell_2 - \ell_1) \phi]
  \} \, .  
\end{equation}
For the ``odd'' part, the diagonal contributions vanish, and one has
\begin{equation}
  W_{|\Psi \rangle}^{(-)} (\ell, \phi)  =  
  \frac{1}{\pi^2} \cos[\phi_0 + (\ell_2 - \ell_1) \phi] \,
  \frac{(-1)^{\ell+(\ell_1+\ell_2 -1)/2}}{\ell_1 +\ell_2 - 2\ell} 
  \delta_{\ell_1 + \ell_2 = \mathrm{odd}} \, ,
  \label{eq:ExamplesSuperOdd2}
\end{equation}
where $\delta_{\ell_1 + \ell_2 = \mathrm{odd}}$  indicates that the sum 
is nonzero only when $\ell_1 + \ell_2$ is odd.

In consequence, when $| \ell_1 - \ell_2|$ is odd, the interference
term contains contributions for any $\ell$, damped as $1/\ell$.  
When $|\ell_1 - \ell_2|$ is an even number, the
contribution~(\ref{eq:ExamplesSuperOdd2}) vanishes and we have 
three contributions: two flat slices coming from the states 
$| \ell_1\rangle$ and $| \ell_2\rangle$ and an interference term 
located at $\ell = (\ell_1 + \ell_2)/2$.

\begin{figure}
 \centerline{\includegraphics[width=0.90\columnwidth]{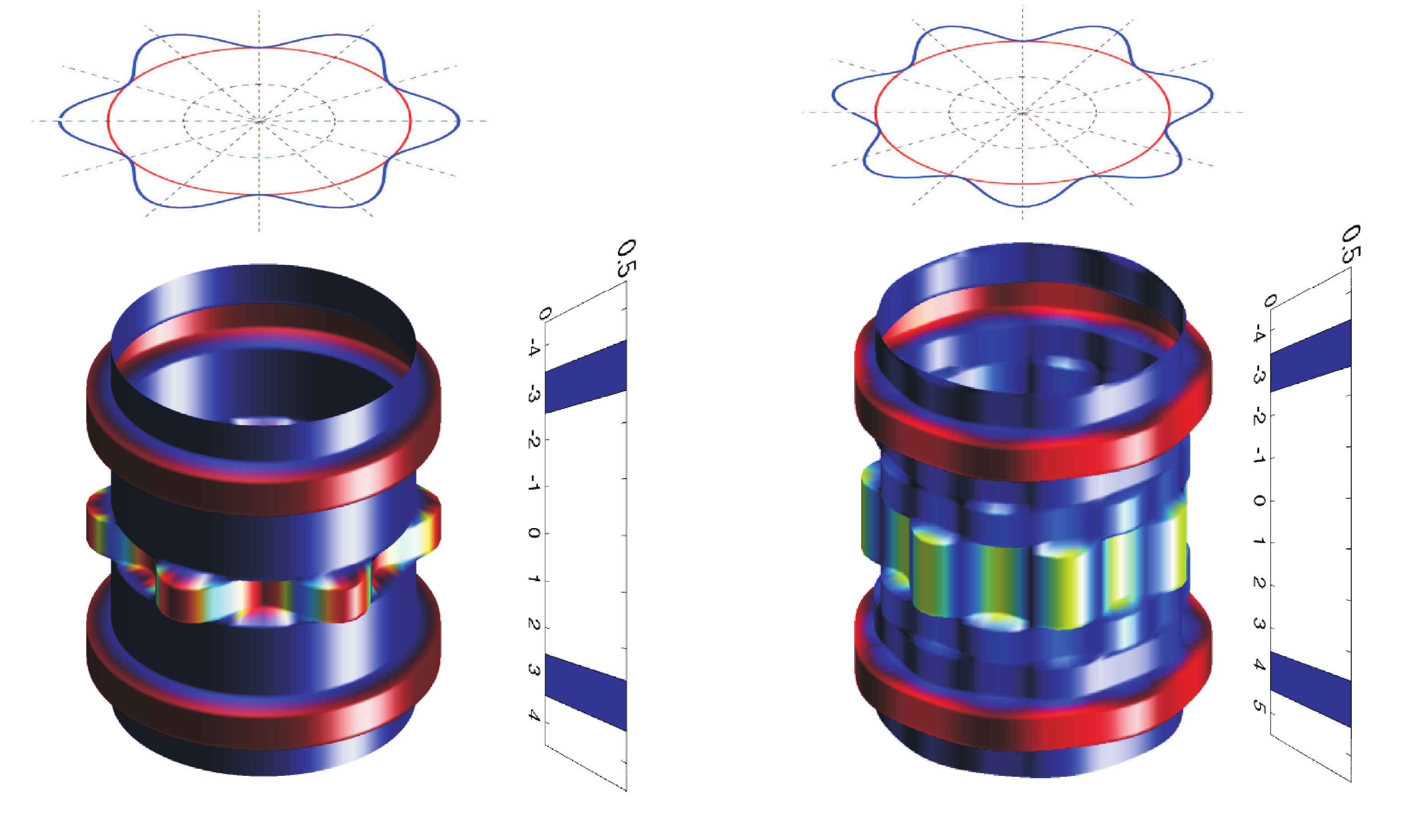}}
    \caption{Plot and marginal distributions
of the Wigner function for an even  superposition 
$|\ell_1 \rangle + e^{i \theta} | \ell_2\rangle $ with $\ell_{1,2}=\pm 3$ 
for  $\ell = -4 $ to $\ell = +4$ (left) and $\ell_1= 4, \ell_2 = -3$ 
for  $\ell = -4 $ to $\ell = +5$.}
\label{fig:EvenTornillo}
\end{figure}

These features are nicely illustrated in figure~\ref{fig:EvenTornillo}. The
state $|\Psi \rangle $ is plotted for $\ell_2 = -3$ and $\ell_1 = 3$
and and $\ell_2 = -3$ and $\ell_1 = 4$.  Changing the relative phase
$\phi_0$ results in a global rotation of the cylinder.  In can be
observed in that the two rings at $\ell = -3$ and $\ell=4$ (as opposed
to the rings at $\ell = \pm 3$), are not flat in $\phi$, but show a
weak dependence on the angle due to the odd contributions added to the
flat Kronecker deltas.

In view of these results, one can wonder what are the pure state for
which the Wigner function is nonnegative.  The answer was found quite
recently~\cite{Rigas:2009rt}: the Wigner function $W_{|\Psi\rangle}
(\ell,\phi)$ is nonnegative if and only if $|\Psi \rangle$ is an
angular momentum eigenstate $|\ell_0\rangle$. The proof is rather
technical and we skip it. The important point we wish to stress is
that while for the continuous case the notions of coherent states,
Gaussian wave packets, and states with nonnegative Wigner functions
are completely equivalent, special care must be paid in extending
these ideas to other physical systems, like angle-angular momentum,
since they lose their equivalence.

To round up this section, we illustrate the benefits of our
phase-space approach by dealing with the evolution of the ``quantum''
pendulum 
\begin{equation}
  \label{eq:PendulumHamilton}
  \op{H} = \frac{\op{L}^{2}}{2}  + \frac{\lambda}{2}
 \left ( \op{E} +  \op{E}^\dag    \right ) \, .
\end{equation}
which has been proposed as a good candidate to describe the evolution
of the wave function of Josephson junctions~\cite{Anglin:2001lq}.  
The corresponding Weyl symbol is 
\begin{equation}
  \label{eq:Symbol}
  h (\ell, \phi ) = \frac{1}{2 \pi} \left ( \frac{\ell^{2}}{2} + \lambda
      \cos \phi \right ) \, .
\end{equation}
 Using the relations \eqref{eq:correspondence}, we get the evolution
 equation for the Wigner function 
 \begin{equation}
   \label{eq:WevolutionExact}
   \frac{\partial}{\partial t} W (\ell, \phi ) = - \frac{1}{2\pi}
   \left  [
     \frac{\ell}{2}  \frac{\partial}{\partial \phi } - 
     2 \lambda \sin \phi \sinh (\delta_\ell /2) 
 \right ]  W (\ell, \phi ) \, . 
\end{equation}
We can directly check that $\exp ( \delta_{ell}/2) W( \ell, \phi) = W
(\ell + 1/2, \phi)$, so (\ref{eq:WevolutionExact})  becomes
 \begin{equation}
   \label{eq:WevolutionExact2}
   \frac{\partial}{\partial t} W (\ell, \phi ) = - \frac{1}{2\pi}
   \left  (
     \frac{\ell}{2}  \frac{\partial}{\partial \phi } W (\ell, \phi)  - 
     \lambda \sin \phi [ W(\ell + 1/2, \phi ) - W ( \ell - 1/2, \phi)     ]
 \right ) \, . 
\end{equation}

We are interested in the semiclassical evolution, i.e., for states
whose angular momentum components are sufficiently concentrated around
a certain value $\ell_0$ with $|\ell_0 | \gg 1$. Thus, we may view
$W(\ell + 1/2, \phi ) - W ( \ell - 1/2, \phi) $ as a derivative
respect to $\ell$, so we finally get
 \begin{equation}
   \label{eq:WevolutionExact3}
   \frac{\partial}{\partial t} W (\ell, \phi ) = - \frac{1}{2\pi}
   \left  (
     \frac{\ell}{2}  \frac{\partial}{\partial \phi }  - 
     \lambda \sin \phi \frac{\partial}{\partial \ell }  \right ) W(\ell, \phi ) \, , 
\end{equation}
whose formal solution is $W (\ell(t), \phi (t))$ in terms  of the
``classical'' trajectories
\begin{equation}
  \label{eq:6}
 \dot{\ell} =- 2 \lambda \sin \phi , 
\qquad 
\dot{\phi}= - \ell .
\end{equation}
This constitutes a nice solution for an involved problem.

\subsection{Tomography}

To complete our theory we propose a reconstruction scheme for
observables in this cylindrical phase space. The equivalent version of
(\ref{eq:Dtom}) reads now, particularized for the density operator,
\begin{equation}
  \label{eq:Wig9}
  \op{\varrho}  =  \suma{\ell} \int_{- \pi}^{\pi}
  \varrho (\ell, \phi) \, \op{D} (\ell, \phi ) \, d\phi \, ,
\end{equation}
where $\varrho (\ell, \phi) = \Tr [ \op{\varrho} \, \op{D}^\dagger
(\ell , \phi) ]/(2\pi)$. In terms of $\varrho (\ell, \phi)$, the
Wigner function is
\begin{equation}
  \label{eq:Wig11}
  W_{\op{\varrho}} (\ell, \phi) = \frac{1}{2\pi}  \suma{\ell^\prime} \int_{- \pi}^{\pi}  
  \exp[ i (\ell^\prime \phi  - \ell \phi^\prime)] \, 
  \varrho (\ell^\prime, \phi^\prime) \, d\phi^\prime \, .
\end{equation}
A reconstruction of this Wigner function is thus tantamount to finding
the coefficients $\varrho (\ell,\phi)$.  For $\ell = 0$, these
coefficients
\begin{equation}
  \label{eq:Tom1}
  \varrho (0, \phi) = \frac{1}{2\pi}  \suma{\ell} \exp( i \ell \phi) \,
  \langle \ell | \op{\varrho} | \ell \rangle 
\end{equation}
are simply the Fourier transform of the angular-momentum
spectrum.

For $\ell \neq 0$, we assume that we are able to experimentally
determine $\op{L}^{2}$ transformations on the input state followed by
angular projections, that is,
\begin{equation}
  \label{eq:Tom7}
  p ( \phi, \zeta)  =
  \langle \phi | \exp( i \zeta \op{L}^{2}/2) \,  \op{\varrho} \,
  \exp(- i \zeta \op{L}^{2}  | \phi \rangle \, .
\end{equation}
One can check that
\begin{equation}
  \label{eq:Tom14}
  \varrho (\ell, \phi) = \frac{1}{2 \pi} 
  \int_{-\pi}^{\pi}   \exp(-i \ell \phi^\prime) \,
  p (\phi^\prime,\phi/\ell ) \, d\phi^\prime \, ,
\end{equation}
so the measurement of $p (\phi, \zeta)$ allows for the determination
$\varrho (\ell, \phi)$ and hence the full reconstruction of the Wigner
function via equation~(\ref{eq:Wig11}).

As a rather simple yet illustrative example, let us note that for the
vortex state $| \ell_0 \rangle$, $\op{L}^{2}$ is diagonal, so the
tomograms $p (\phi^\prime, \phi^\prime/\ell)$ are independent of
$\phi$ and $\ell$ and all of them equal to $1/(2\pi)$.  Performing the
integration we obtain precisely the Wigner function in equation
(\ref{eq:ExampleOAMstaleEll}).

The feasibility of the proposed scheme relies on two crucial points.
First, the implementation of the $\op{L}^{2}$ transformation, which
corresponds to a free rotor. This has been used to describe the
evolution in a variety of situations. Second, we need to assess the
measurement of the angular spectrum, which can be done only
approximately.  Though the implementation of this scheme may differ
depending on the system under considerations, our formulation provides
a common theoretical framework on the Hilbert space generated by the
action of angle and angular momentum. A experimental demonstration of
the method in terms of optical beams has been recently
established~\cite{Rehacek:2010fk}.

For continuous variables, it is possible to make a direct sampling of
the Wigner function by a scheme that determines the
parity~\cite{Englert:1993nx,Bertet:2002rr}. This is an unexplored
territory for the cylinder that is worth investigating.

\section{Concluding remarks}

In summary, we have shown how to extend in a consistent way all
the techniques developed for a continuous-variable phase space to the case
of angle and angular momentum. While we have not left aside the
mathematical details, our main emphasis has been on presenting a
simple and useful toolkit that any practitioner in the field should
master. In our view, far from being an academic curiosity, the ideas
expressed here have a wide range of potential applications in
numerous hot topics.

This paper has greatly benefited from the criticism and advice of
Prof. B.-G. Englert. The work was supported by the Spanish Research
Directorate (Grants FIS2005-06714 and FIS2008-04356), the UCM-BCSH
program (Grant GR-920992), the Mexican CONACyT (Grant 106525), the
Czech Ministry of Education (Project MSM6198959213), and the Czech
Ministry of Industry and Trade (Project FR-TI1/364).


\end{document}